\documentclass[5p,twocolumn]{elsarticle}
\journal{Computer Networks}
\usepackage{amsmath}
\usepackage{amsfonts}
\usepackage{amssymb}
\usepackage{graphicx}
\usepackage[ruled]{algorithm2e} % Builds nice algorithms.

\newtheorem{theorem}{Theorem}
\newtheorem{corollary}[theorem]{Corollary}
\newtheorem{definition}{Definition}
\newtheorem{example}{Example}
\newtheorem{Rule}{Rule}
\newproof{proof}{Proof}

\begin{document}

\begin{frontmatter}

\title{Chain Routing: A new routing framework for the Internet based on complete orders}

\author[UoB]{P. David Arjona-Villica\~{n}a\corref{cor1}}
\ead{darjona@yahoo.com}

\author[UoB]{Costas C. Constantinou}
\ead{c.constantinou@bham.ac.uk}

\author[Aston]{Alexander S. Stepanenko}
\ead{a.stepanenko@aston.ac.uk}

\address[UoB]{University of Birmingham, Edgbaston, Birmingham, B15 2TT, United Kingdom}
\address[Aston]{Aston University, Aston Triangle, Birmingham, B4 7ET, United Kingdom}

\cortext[cor1]{Corresponding author}

%---------------------------------------------------------------
\begin{abstract}
A new framework to perform routing at the Autonomous System level is proposed in this paper. This mechanism, called Chain Routing, uses complete orders as its main topological unit. Since complete orders are acyclic digraphs that possess a known topology, it is possible to define an acyclic structure to route packets between a group of Autonomous Systems. The adoption of complete orders also allows easy identification and avoidance of persistent route oscillations, eliminates the possibility of developing transient loops in paths, and provides a structure that facilitates the implementation of traffic engineering. Moreover, by combining Chain Routing with other mechanisms that implement complete orders in time, we suggest that it is possible to design a new routing protocol which could be more reliable and stable than BGP's current implementation. Although Chain Routing will require an increase of the message overhead and greater coordination between network administrators, the rewards in stability and resilience should more than compensate for this effort.
\end{abstract}


\end{frontmatter}

%---------------------------------------------------------------
\section{Introduction}\label{intro}

The Border Gateway Protocol (BGP) has been the Internet's \emph{de facto} routing protocol at the Autonomous System (AS) level since its deployment in 1993. Currently defined in RFC 4271 \cite{Rekhter2006}, BGP is a decentralized routing algorithm in which every router independently computes its best path to each destination in its routing table. When a group of ASs and its routers adopt a stable set of paths to reach a destination, it is said that the network \emph{converges} or finds a \emph{solution}. However, BGP has been deemed to be unstable because it is prone to develop persistent cyclic behavior \cite{Varadhan1996}, and to suffer from delays during convergence \cite{Labovitz2000, Mao2002}.

BGP routers form a complex distributed routing system with a rich set of interactions. In some situations, these may cause an excesive number of messages which consequently result in delays in the convergence of a BGP network \cite{Labovitz1999, Labovitz2000, Mao2002}. This pathology is usually triggered by changes in the network's topology. Different solutions have been proposed to attain shorter convergence times, and the most recent ones \cite{Chandrashekar2005, Pei2005} attempt to define a temporal order to limit the undesirable effects of excesive messaging. Unfortunately, these mechanisms do not take into consideration the persistent route oscillations (PRO) that may develop in a BGP system \cite{Varadhan1996}. Therefore, another independent strand of research has tried to solve this specific problem \cite{Griffin1999, Griffin2002, Gao2000a, Cobb2004, Ee2007}. However we argue that before any major modifications or replacements can be made to BGP, it is first necessary to analyze the topological structure of the network in which this routing protocol is employed. In other words, what would be the best routing framework to deliver information in a network with structure and topology similar to the Internet's?

In an earlier publication \cite{Arjona-Villicana2010}, we demonstrated that, in its network core, the European section of the Internet possesses rich path diversity which BGP does not currently exploit. A routing protocol which could exploit the Internet's path diversity may increase this network's resilience to failures and allow an effective implementation of traffic engineering. Evidently, an increase in the number of paths available may translate in a more complex routing algorithm, but this could be a fair price to pay for greater resilience and better traffic managment, provided the overheads are quantifiable and bounded.

In this paper we address BGP's inherent instabilities and its inability to exploit the Internet's path diversity by proposing a new routing framework, which we call Chain Routing. This framework employs a new topological unit, the complete order, to define acyclic paths to a destination. Consequently, we will try to demonstrate that Chain Routing could help increase the Internet's resilience to failures and employ its path diversity. We go as far as demonstrating that our proposal is a feasable idea which could be implemented as one of the main components of a routing protocol; however we do not define a fully-fledged routing protocol. Moreover, concrete proof that this framework performs better than BGP's current implementation is not offered here, but forms part of an ongoing research which we aim to pursue in the future.

This paper is organized as follows. Section \ref{background} provides the background to this research. The mathematical concepts needed to justify the framework proposed here are introduced in Section \ref{comp_orders}. Then Section \ref{chain_rtg} provides a description of how complete orders can be employed to perform routing in a network and a demonstration that this model is feasable for implementation, via a small numerical analysis. In Section \ref{applying} an evaluation of the potential application of Chain Routing to increase network stability and its implementation costs are presented. Section \ref{comp_orders_in_time} demonstrates how complete orders in time might be also neeeded to enhance the stability of a network. Finally, Section \ref{discusion} discusses the advantages and disadvantages of using this framework before arriving at conclusions in Section \ref{conclusion}.

\section{Background}\label{background}

When a BGP router announces a destination, it also announces the path that is used to reach back this destination. BGP uses Classless Interdomain Routing (CIDR) prefixes as destinations and the list of ASs that have passed the announcement as paths. Therefore, this protocol has been classified as a path-vector routing protocol.

BGP selects only one path to reach back each destination in its routing table \cite{Rekhter2006}. We name this BGP property the \textbf{Preferred Paths Rule} (PPR). It is important to notice that this restriction is a characteristic of the BGP protocol and not of the relationship that exists between ASs in the Internet. Proof of this assertion is the fact that many network administrators try to override the PPR in order to employ more than one route to reach a destination and thus implement traffic engineering \cite{Caesar2005}. It also illustrates that BGP was not originally designed to support traffic engineering.

Another important BGP characteristic is that the selection and announcement of a preferred path  can be modified by the network administrator. These adjustments are needed in order to accomodate the comercial agreements made by the owners of each AS. This BGP property is known as policy configuration, or just \emph{policies}, and it plays an important role in the functionality of any routing protocol at the AS level.

The Internet's topology is usually modelled as an undirected graph, $G=(V,E)$, in which ASs are represented by the vertex set $V$, and their communication links by the edges $E$ that join them. Furthermore, in \cite{Arjona-Villicana2010} we proposed to use digraphs, $D=(V,A)$, to represent how the announcement of destinations is restricted by policies. Consequently, directed edges or arcs $A$ would be used to model how the policies, implemented at ASs, shape the propagation of destinations in a network. This graph was called the \emph{announcement digraph of destination i} $D_{anc}(i)$, and its converse, the \emph{destination digraph of i} $D_{dst}(i)$, could represent how ASs may use different paths to reach back the announced destination.

On the other hand, the restrictions imposed by the PPR mean that, after selecting its preferred path, an AS can only announce one route per destination to its neighboring ASs. Therefore, destinations in the Internet are announced following an \emph{oriented tree} or \emph{arborescence}, which we call the \emph{BGP digraph of destination i} $D_{BGP}(i)$. Nevertheless, the BGP digraph does not show how destinations could propagate through the Internet if only the restrictions imposed by policies are taken in consideration (i.e., BGP's PPR is ignored).

The graph-theoretic representations we propose above allow a better understanding of how information flows and how instabilities originate and cause problems in the Internet. We have divided the BGP instabilities observed by other authors in two different categories: persistent route oscillations (PRO) and convergence delays.

\subsection{Persistent route oscillations}\label{pro}

An early study on the instabilities of the Internet \cite{Varadhan1996} uncovered that BGP could develop PRO when a group of ASs cannot find a unique solution to reach a destination due to conflicting BGP policies. This causes BGP routers to fall in a state where they alternate between different paths to a destination in an endless cyclic behavior.

Further analysis of this pathology \cite{Griffin1999, Griffin2002} demonstrated that, in order to avoid the development of this problem, it is necessary to eliminate the cyclic relationships that may exist between the ASs involved in the selection process. To achieve this objective a mathematical model called a \emph{dispute wheel}, was proposed by Griffin et al.  \cite{Griffin1999, Griffin2002}, who demonstrated that when dispute wheels do not develop, the network is free of PRO. An alternative way to formulate the dispute wheel model is that it tries to avoid the development of cycles in the announcement digraph of a destination

A different approach to avoid PRO was obtained through a set of guidelines which restrict the paths an AS can use to reach its neighbors depending on the commercial relationship between them \cite{Gao2000a}. Such guidelines reinforce the Internet's hierarchical structure and eliminate some of the potentially problematic alternative paths. Closer inspection of this model shows that it employs policies to restrict the network's path diversity and to adopt a directed tree when announcing and reaching back destinations.

Recent solutions to the PRO problem \cite{Cobb2004, Ee2007} propose to use different metrics to determine when a BGP system is oscillating and stop this behavior. Unfortunately, a common requisite of all the mechanisms described in this section is that BGP will always enforce its PPR to find a unique best path for each destination, which in turn fails to exploit the path diversity available at the core of the Internet \cite{Arjona-Villicana2010}. We believe this is a missed opportunity since path diversity could be employed to increase the capacity and resilience of the network.

\subsection{Convergence delays}\label{conv_delays}

BGP's slow convergence is said to be caused by the excessive message exchange that sometimes develops after the network's topology has changed, while the error messages propagate through the network and until a new solution is reached; some authors have called this transient state \emph{path exploration} \cite{Chandrashekar2005}. There have also been proposals to speed up the convergence of BGP networks \cite{Pei2002, Bremler-Barr2003}. The reason why so many resources have been devoted to study this problem is because, when the network is in a state of path exploration, it is more vulnerable to develop transient loops and these could in turn, cause packets to be dropped.

Two of the most recent solutions to this problem \cite{Pei2005, Chandrashekar2005} advocate using timestamps which effectively implement an absolute temporal order of the control messages that are produced in a network. Later, in Section \ref{comp_orders_in_time}, we will return to this mechanisms because we believe that time is an important constraint in the correct functionality of a routing protocol.

The previous solutions to the convergence delays experienced in the Internet assume that the network will finally converge, but they fail to consider what could happen if the system develops PRO. As other researchers have demonstrated (Section \ref{pro}), in order to avoid PRO it is imperative to eliminate the directed cycles present in the announcement and destination digraphs. The literature on the field of acyclic digraphs \cite{Harary1965, Bang-Jensen2002} proves that the maximal acyclic digraph is the \emph{complete order}. The following section provides the mathematical background needed to use topological complete orders in a graph. Later we will also discuss and apply complete orders in time.

\section{Complete Orders}\label{comp_orders}
In contrast to (undirected) edges, \emph{arcs} possess a direction, this means that an arc labeled $ab$ has a \emph{tail} ($a$) and a head ($b$) which represent the direction of the arc. An arc with the same head and tail ($aa$) is called a \emph{loop}. And a \emph{cycle} is a closed directed path of two or more arcs.

\begin{figure}[!t]
	\centering
	\includegraphics[width=3in]{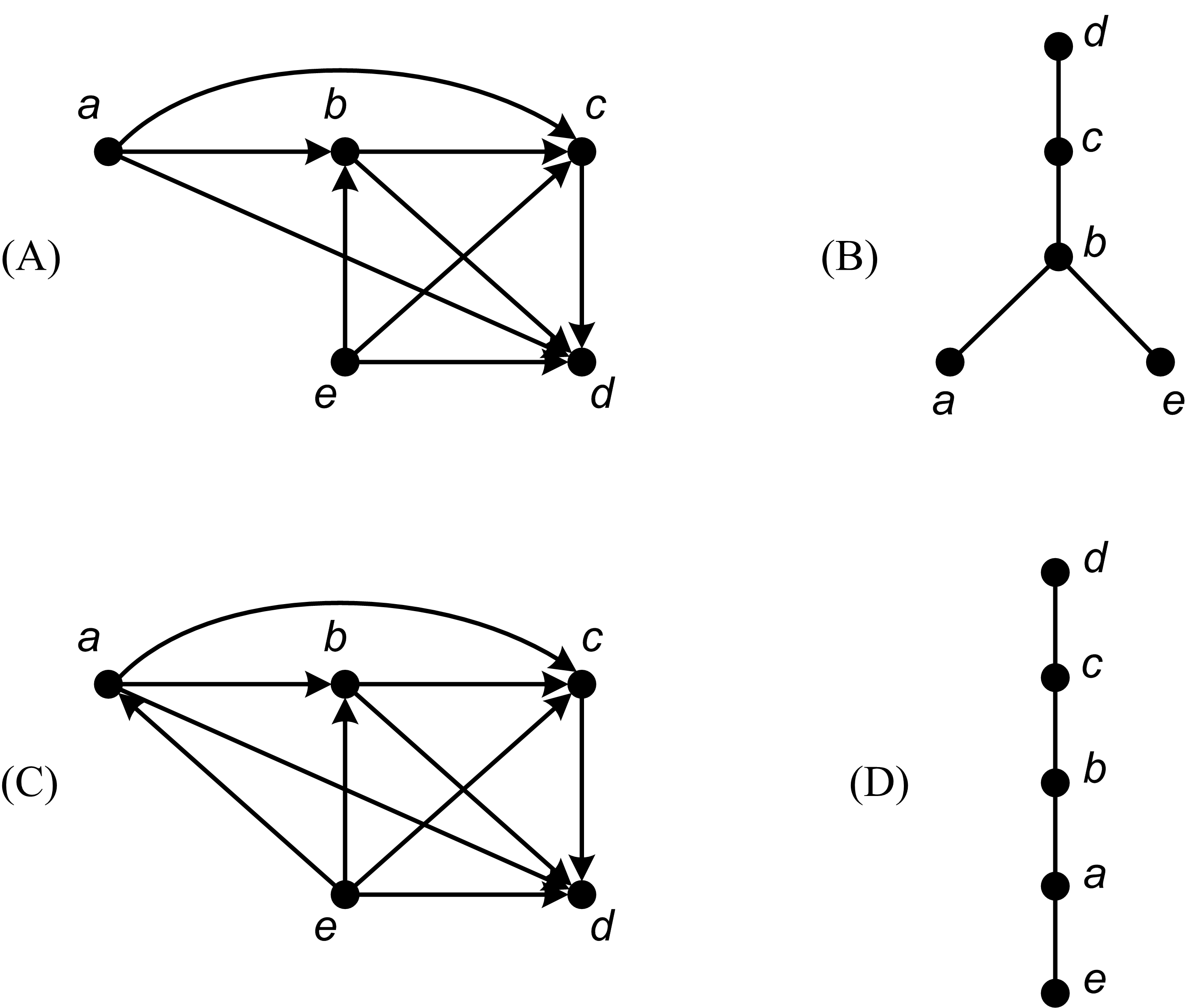}
	\caption{A partial order (A) and its Hasse diagram (B); a complete order (C) and its Hasse diagram (D)}
	\label{fig:Hassediagram}
\end{figure}

Any digraph with no cycles is called an \emph{acyclic digraph}. A \emph{partial order} is an acyclic digraph in which the vertices possess the following three properties: \emph{irreflexive} (there are no loops), \emph{asymmetric} (if arc $ab$ exists, then arc $ba$ cannot exist) and \emph{transitive} (if arc $ab$ and $bc$ exist, then arc $ac$ must exist). When a partial order is also \emph{complete}, that is, all the previous properties apply to all the vertices in the digraph, then it is called a \textbf{complete order}, \emph{total order} or \emph{linear order}. Examples of both types of orders are provided in Fig.\ \ref{fig:Hassediagram}. It is known that a complete order with $n$ vertices has $n(n-1)/2$ arcs \cite{Harary1965}.

Two digraphs are \emph{isomorphic} when there is a one-to-one correspondence between their vertices and their arcs. Because of their completeness, all complete orders with the same number of vertices are isomorphic. It is also said that complete orders are \emph{maximal}, because adding a new arc to this digraph forms a cycle. Therefore, a complete order is the maximal acyclic digraph that can be formed given a set of vertices. Another important property of complete orders is:

\begin{theorem}[from \cite{Harary1965}]
Every complete order has a unique transmitter and a unique receiver.
\end{theorem}

This means that the subgraph of a destination digraph with the largest possible number of arcs, which is still acyclic and thus provides the greatest path diversity between a source and a destination, is a complete order.

A digraph can also be described as a binary relation on its vertex set, where each of the existing arcs $ab$ in the digraph is equivalent to the binary relation $ aRb$. Therefore, properties for binary relations can be applied to digraphs and \emph{vice versa}.

Besides digraphs, another common representation of partial orders is the \emph{Hasse diagram}: To construct this from a digraph $D$, first draw the vertices of $D$ in vertical order such that $a$ is below $b$ if $aRb$, then draw all the digraph's arcs (which should have an upward direction if the order was done correctly), delete all arcs that could be implied by the transitive property, and finally delete the direction indicators in the remaining arcs. Examples of Hasse diagrams for a partial and a complete order are shown in Fig.~ \ref{fig:Hassediagram}.

A common mathematical notation of a partial order on a set $S$ is $(S,\prec)$ \cite{Fishburn1985}. A complete ordered subset of a partial order is also called a \textbf{chain}. For example, in Fig.~ \ref{fig:Hassediagram}(B) vertices $a$, $b$, $c$ and $d$ form a chain, denoted here as $C(a, b, c, d)$. The \emph{height} $H(S,\prec)$ of a partial order $(S,\prec)$ is one less than the number of vertices in a maximum length chain in $(S,\prec)$. Therefore, the partial order in Fig.~ \ref{fig:Hassediagram}(B) has height 3, because both maximum length chains, $C(a, b, c, d)$ and $C(e, b, c, d)$, have 4 vertices; and the complete order's height is 4.

In a previous publication \cite{Arjona-Villicana2010} we proposed to use the number of arc-disjoint paths as a metric of \emph{path diversity} between a source and a destination.  A group of \textbf{arc-disjoint paths} is a set of paths connecting two vertices through intermediate vertices, in which none of the paths traverse the same arc more than once. Notice that arc-disjoint paths can visit the same vertex more than once, as long as different arcs are used to reach and leave each vertex. A group of arc-disjoint paths is a resilient strategy to send information from a source to a destination because paths do not share arcs or communication links. The number of arc-disjoint paths in a complete order is defined by the following theorem:

\begin{theorem}\label{T:Num2}
If $C$ is a complete order with $n\geq2$ vertices, then there are $n-1$ arc-disjoint paths from the transmitter to the receiver.
\end{theorem}

\begin{proof}
This is proven by induction:

The smallest possible 2-vertex complete order has only 1 path: the arc $v_{1}v_2$. Now assume that a complete order with $n$ vertices has $n-1$ arc-disjoint paths between the transmitter $v_1$ and the receiver $v_n$. Then, the transitive property implies that the complete order with $n+1$ vertices also has $n-1$ arc-disjoint paths between the transmitter $v_1$ and its receiver $v_{n+1}$. In addition to these, the arc $v_{1}v_{n+1}$ provides an additional arc-disjoint path, because of the completeness property. This results in a total of $n$ arc-disjoint paths from $v_1$ to $v_{n+1}$.
\end{proof}

Notice that the number of arc-disjoint paths and the height of a complete order $H(C,\prec)$ are the same. Theorem \ref{T:Num2} also demonstrates that the smallest complete order that offers any path diversity has 3 vertices, and thus, just 2 arc-disjoint paths. Still, it is necessary to determine how many arcs does the  $n-1$ arc-disjoint paths from the transmiter to the receiver will use ($u$) in a complete order, and how many arcs will remain unused ($r$) by these paths:

\begin{theorem}\label{T:Num3}
If $C$ is a complete order with $n\geq2$ vertices, then all of the $n-1$ arc-disjoint paths from the transmitter to the receiver use exactly $u=2n-3$ arcs.
\end{theorem}

\begin{proof}
This is also proven by induction:

The smallest possible 2-vertex complete order uses only 1 arc for its only path: the arc $v_{1}v_2$. Now assume that a complete order with $n$ vertices uses $2n-3$ arcs on its $n-1$ arc-disjoint paths between the transmitter $v_1$ and the receiver $v_n$. Most of these arcs follow a path from the transmitter to an intermediate node and then to the receiver, and there is just 1 direct path from the transmiter to the receiver. Then, because of its transitivity, the complete order with $n+1$ vertices must also use $2n-3$ arcs on its arc-disjoint paths between the transmitter $v_1$ and the receiver $v_{n+1}$ plus the arc $v_{n}v_{n+1}$ and, because of its completeness, the arc $v_{1}v_{n+1}$. This gives a total of:

\[
u = 2n-3+1+1 = 2(n+1)-3.
\]
\end{proof}

\begin{corollary}\label{T:Num4}
If $C$ is a complete order with $n\geq2$ vertices, there are exactly $r={(n-2)(n-3)}/2$ arcs that are not used in all the $n-1$ arc-disjoint paths from the transmiter to the receiver.
\end{corollary}

\begin{proof}
This result is easily obtained by substracting the number of arcs used by the $n-1$ arc-disjoint paths (Theorem \ref{T:Num3}) from the complete order's total number of arcs: $n(n-1)/2$.
\end{proof}

Notice that $r$ increases quadratically with respect to the number of vertices in the complete order ($n$), while $u$ only does so linearly. By closer inspection it is possible to see that when $n=3$, $u=3$ and $r=0$, but because $r$ increases quadratically, when $n=8$, $u=13$ and $r=15$. This means that when $n>7$, $r$ becomes larger than $u$. This may translate into ASs and their routers spending more resources to store and manage the $r$ stand-by arcs, instead of the $u$ arcs which form the primary arc-disjoint paths. Therefore, in practice, complete orders that grow beyond a \textbf{maximum size} of 7 vertices could be too costly to be considered a sensible solution for a communication network's routing needs.

Another important characteristic of complete orders is that they offer flexibility and predictability when changing their height (i.e., number of vertices). The following theorem demonstrates how easy it is to reduce the height of a complete order:

\begin{theorem}[from \cite{Harary1965}]\label{T:Num5}
If $C$ is a complete order with at least 3 vertices, and if $v$ is any vertex of $C$, then $C-v$ is also a complete order.
\end{theorem}

This previous theorem demonstrates that, when a chain of $n$ vertices needs to eliminate a vertex, it is possible to end with the chain of $n-1$ vertices. Conversely, the next theorem and corollary analyze how many arcs will be needed to increase the height of a complete order:

\begin{theorem}\label{T:Num6}
If $C$ is a complete order with $n\geq2$ vertices, then the augmented complete order $C+1$ which has exactly 1 more vertex than $C$ requires $n$ new arcs.
\end{theorem}

\begin{proof}
$C$ has $\dfrac{n(n-1)}{2}$ arcs. So $C+1$ has $\dfrac{(n+1)n}{2}$ arcs.

Hence the difference in the number of arcs between these complete orders is:
\[
\frac{n^2+n}{2} - \frac{n^2-n}{2} = n.
\]
\end{proof}

\begin{corollary}\label{T:Num7}
Increasing the number of vertices and the height of a complete order by 1 requires at least 2 new arcs.
\end{corollary}

\begin{proof}
Since the smallest possible complete order has 2 vertices, then 2 new arcs are needed to augment to the complete order of 3 vertices.
\end{proof}

Theorem \ref{T:Num6} and Corollary \ref{T:Num7} demonstrate why a single arc is not enough to increase the height of a chain.

\section{The Chain Routing framework}\label{chain_rtg}

We propose Chain Routing as a routing framework that employs complete orders (or chains) as the basis for determining a set of valid routes to a destination. Such a framework has two main advantages:

\begin{enumerate}
	\item It uses the maximum number of acyclic directed paths between two nodes (vertices) as its natural unit of path diversity.
	\item It requires a simple data structure to store several paths (Section \ref{crds}).
\end{enumerate}

The main objective of Chain Routing is to define a chain between the source $s$ and a destination $d$ which includes as many intermediate vertices as possible, provided the chain's maximum allowed size of 7 vertices is maintained. Vertices included in this set, other than $s$ and $d$, are called the \emph{intermediate nodes} from $s$ to $d$. Because of the completeness and maximality properties of complete orders, a chain can be used to represent a \emph{self-contained strategy} to reach $d$, even when some of the intermediate nodes or links fail. This means that after a chain has been defined, all the ASs in the chain will transmit information following the same paths described by the topology of the complete order.

Since Chain Routing can be thought either as a replacement or as an enhancement to BGP, it can operate as a decentralized routing algorithm. This means that each AS must learn the network's topology through the announcements received from its neighbors, and also that each AS will need to define its own set of chains to different destinations. However, contrary to BGP, more coordination between ASs is needed to define the chains and to set the order of its intermediate nodes. The following example shows how some of the properties of chains could apply to destination digraphs:

\begin{example}
Assume Fig.\ \ref{fig:Hassediagram}(C) represents a destination digraph for AS $d$, $D_{dst}(d)$. There are many routes $e$ may choose to send packets to $d$, but the following 4 arc-disjoint paths provide the most resilient strategy:

\begin{enumerate}
	\item $e \rightarrow d$
	\item $e \rightarrow a \rightarrow d$
	\item $e \rightarrow b \rightarrow d$
	\item $e \rightarrow c \rightarrow d$
\end{enumerate}

There are also 4 not arc-disjoint paths that $e$ could use to reach $d$:

\begin{enumerate}
	\item $e \rightarrow a \rightarrow b \rightarrow d$
	\item $e \rightarrow a \rightarrow c \rightarrow d$
	\item $e \rightarrow b \rightarrow c \rightarrow d$
	\item $e \rightarrow a \rightarrow b \rightarrow c \rightarrow d$
\end{enumerate}

When $e$ follows the proposed strategy and uses arc-disjoint paths, it could balance the traffic load between each of these paths, or it could prefer to use the direct path ($e \rightarrow d$) and leave the other arc-disjoint paths as backup. If $e$ picks the latter option, and later link $ed$ fails, $e$ has still 3 safe alternative paths to route to $d$.

Regardless of the individual computations that each AS will need perform in order to determine its preferred chain to a destinaiton, each chain will still need coordination between the source and the intermediate nodes in order to avoid instabilities in this system. For example, if $e$ distributes its traffic between the 4 arc-disjoint paths, each of the intermediate nodes will need to know the strategy followed by $e$, otherwise $a$ could use a non arc-disjoint path, like $e \rightarrow a \rightarrow b \rightarrow d$, which may cause congestion with path $e \rightarrow b \rightarrow d$.
\end{example}

Notice that none of the paths in the previous example can form a cycle, even if two or more paths are combined. This is because the original destination digraph is acyclic. Therefore it is possible to guarantee that, regardless of which vertex or arc fails, cyclic paths will not develop in this network. In other words, when an arc or a vertex (with its adjacent arcs) is removed from an acyclic digraph, the result is still an acyclic digraph.

It is our intention that, order of the ASs in a chain determines how information packets flow in the network, but not how control messages are exchanged between ASs. In other words, ASs could be allowed to break the chain order to quickly communicate a change in the network's topology or the occurrence of a failure which causes loss of connectivity.

\subsection{Chain Routing data structure}\label{crds}

The main objective of this structure is to keep a correspondence between the network's topology and the chains used to route information through it. In order to achive this, the Chain Routing data structure will store three different \textbf{basic structures}, and two of these may also contain (or point) to other basic structures. We will also use the concept of \textbf{levels of abstraction} to represent that a structure may recursively contain other structures. Initially structures will be stored at level 0, and these may contain other structures at level 1, which may also contain other structures at level 2, and this continues until all the structures of the topology have been recorded. The three basic structures we propose for Chain Routing are:

\begin{description}
	\item[arc] The most basic structure only records a link between two ASs and it cannot contain other structures.
	\item[Varc] It describes a path which, in terms of the data structure, is just a sequence of arcs. A Varc usually contains arcs at the next level of abstraction.
%	\item[chain] henceforth we will call \textbf{segments} to the $n(n-1)/2$ chain's arcs, in order to remember that each segment may contain another basic structure a the next level of abstraction.
	\item[chain] A chain or complete order may contain any of the three basic structures at the next level of abstraction; henceforth the arcs of a chain will be called \textbf{segments} to denote that they may be of different type to the basic arc structure.
\end{description}

The Varc structure mentioned before records ASs which do not possess enough connectivity to form a chain, but that still allow transmitting packets through them and hence, could be considered as a sequence of arcs that follow a predefined path:

\begin{definition}\label{D:david_69}
A \textbf{Varc} or virtual arc is a structure that represents a set of vertices and their adjacent arcs that follow a directed path from an initial vertex $x$ to a final vertex $y$, where the directed path needs to be abstracted in order to allow $x$ and $y$ to be the end vertices of a chain segment.
\end{definition}

A new link must be initially recorded in the data structure as an arc at level 0, but it is necessary to consider that this new arc could also:

\begin{enumerate}
	\item allow defining a new or a longer chain.
	\item help to combine two chains, with common vertices, into a longer one.
	\item be included as part of a Varc.
\end{enumerate}

From the three previous options, arcs that can be included as part of a Varc should have the least preference, because Varcs do not increase the path diversity of the network. The other two options should only be preferred depending on which one will form the chain with greater height.

\begin{figure}[!t]
	\centering
	\includegraphics[width=3.1in]{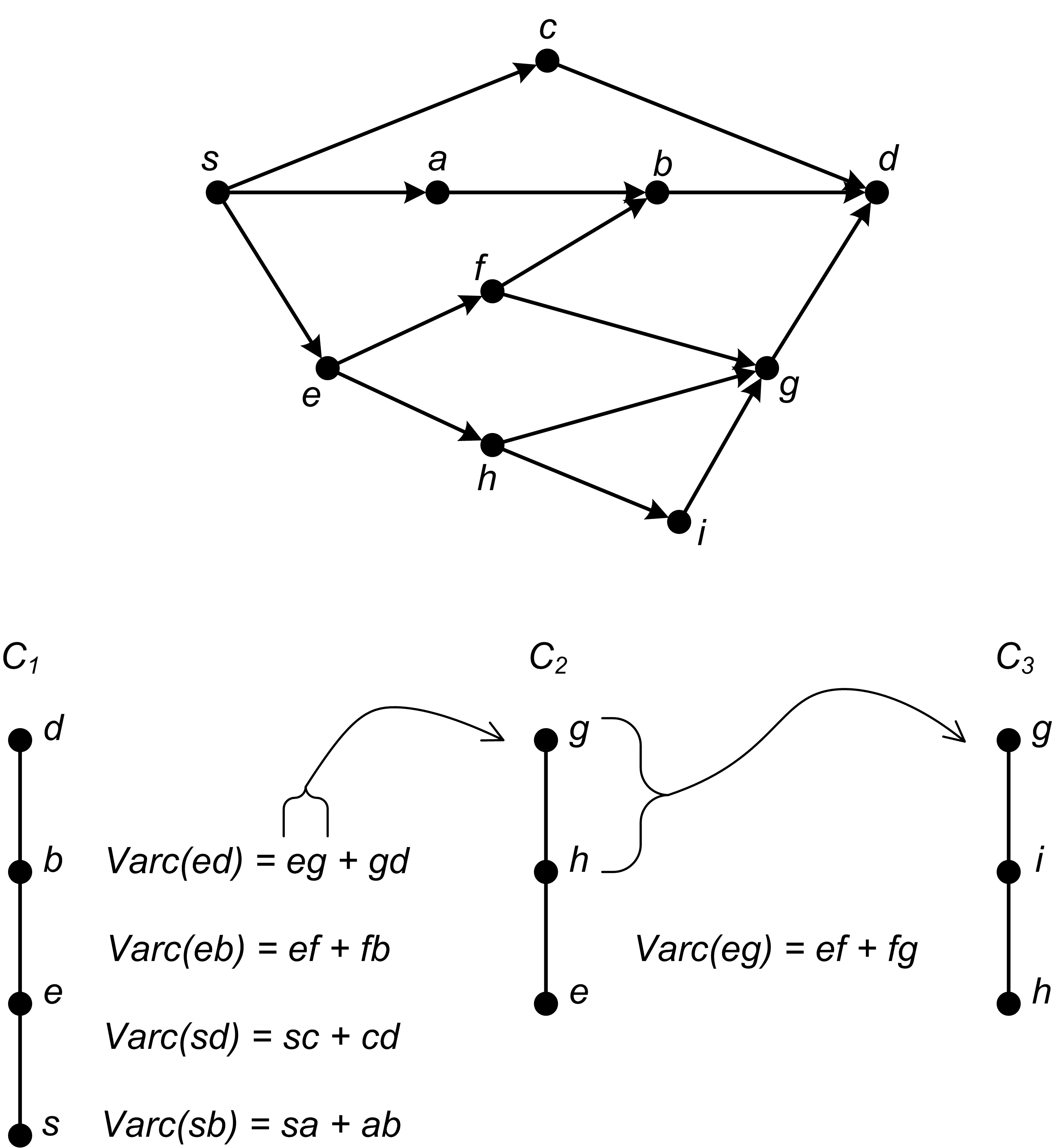}
	\caption{Chain Routing example}
	\label{fig:AlgorithmExample}
\end{figure}

Now we provide an example of how the Chain Routing data structure could be employed to describe the topology of a network:

\begin{example}
The network depicted in Fig.\ \ref{fig:AlgorithmExample} shows a 10-vertex destination digraph which has been represented or abstracted using the 4-vertex chain $C_1(s,e,b,d)$, which must be stored at level 0 in the Chain Routing data structure. Such a chain has 6 segments that need to be recorded at level 1:

\begin{enumerate}
	\item Segment $se$ is an arc.
	\item Segment $sb$ is a Varc formed of arcs $sa$ and $ab$.
	\item Segment $sd$ is a Varc formed of arcs $sc$ and $cd$.
	\item Segment $eb$ is a Varc formed of arcs $ef$ and $fb$.
	\item Segment $ed$ is a Varc fromed of segment $eg$ and arc $gd$.
	\item Segment $bd$ is an arc.
\end{enumerate}

The arcs and segments of each Varc are recorded at the next level of abstraction (2). Segment $eg$ of Varc $ed$ is abstracted as chain $C_2(e,h,g)$, and its segments are stored at level 3:

\begin{enumerate}
	\item Segment $eh$ is an arc.
	\item Segment $eg$ is a Varc formed of arcs $ef$ and $fg$.
	\item Segment $hg$ is a chain $C_3(h,i,g)$.
\end{enumerate}

Finally, each of the arcs of Varc $eg$ and the segments of $C_3$ are stored at level 4 of the data structure.
\end{example}

%An example of how Chain Routing and its data structure could be used to describe the topology of a network is provided in Fig.\ \ref{fig:AlgorithmExample}, which shows how a 10-vertex destination digraph could be represented using a 4-vertex chain. This figure demonstrates how, by nesting chains and Varcs, it is possible to abstract many vertices and arcs in this network. The main chain in this example, $C_1$, uses $e$ and $b$ as its intermediate nodes and it only has a height of 3, which shows that the number of vertices in a chain may be significantly less than the number of nodes in its destination digraph. Nesting structures and virtual arcs allow to condense the network's topology and, more importantly, to focus attention on the ASs which are central to the path diversity and resilience of the network.

The previous example demonstrates how, by nesting chains and Varcs, it is possible to abstract many vertices and arcs in a network. The main chain in this example, $C_1$, uses $e$ and $b$ as its intermediate nodes and it only has a height of 3, which shows that the number of vertices in a chain may be significantly less than the number of nodes in its destination digraph. Nesting structures and virtual arcs allow to condense the network's topology and, more importantly, to focus attention on the ASs which are central to the path diversity and resilience of the network.

In order to define the structures shown in Fig.\ \ref{fig:AlgorithmExample}, a great degree of coordination between the ASs in this network is needed. For example, AS $e$ has four valid options for an intermediate node to $d$: $b$, $f$, $g$ and $h$; but only $b$ allows the definition of $C_1$ as depicted. Therefore, $s$ would need to coordinate with $e$ in order to obtain the solution shown in this figure.

Notice that the previous example is not the only solution which could be used to represent the network in Fig. \ref{fig:AlgorithmExample} using chains, but this paper does not try to provide the final and full implementation of the Chain Routing framework, just to prove that complete orders could be used as a safe method to perform routing.

\subsection{A naive implementation of Chain Routing}\label{experiment}

To demonstrate the applicability of our idea to a directed graph and that is feasible to use complete orders to represent the path diversity of the Internet, we implemented a computational program capable of finding as many chains as possible in the 45 announcement digraphs of European ASs that we obtained in a previous analysis of the Internet \cite{Arjona-Villicana2010, Arjona-Villicana2009}.

We presume that the simplest way to identify chains in a digraph is by using the transitive relationship that exists between its vertices. Therefore, a vertex that can be reached through more than one path must posses a transitive relationship with at least one other vertex in this graph. The program we developed is based in the Breadth-First Search (BFS) algorithm \cite{Bang-Jensen2002}. We decided to use BFS because it discovers first the direct paths from a source vertex, and later the paths that use intermediate nodes. We include here our modified BFS algorithm under Algorithm \ref{alg:main_algorithm}. The input to this program is the adjacency matrix of an announcement digraph.

%To keep track of the vertices that have been visited, this algorithm uses the \emph{distance} array which will either contain a valid value, in which case the vertex has been visited before and a transitive relation exists, or an initial invalid value, which means the vertex is currently being visited for the first time. 

\begin{algorithm}[!t]
	\SetAlgoNoLine
	\KwIn{The adjacency matrix of $D_{anc}(i)=(V,A)$}
	%\KwOut{No output needed from this algorithm}
	\BlankLine

	\ForEach{$v \in V$}{
		predecessor($v$) $\leftarrow$ nil\;
		distance($v$) $\leftarrow$ nil\;
	}
	\BlankLine

	distance($i$) $\leftarrow$ 0\;
	QUEUE $\leftarrow$ $i$\;
	\BlankLine

	\While{QUEUE $\neq \emptyset$}{
		$x$ $\leftarrow$ head of QUEUE (also delete $x$ from QUEUE)\;
		\BlankLine
		\ForEach{out-neighbour ($y$) of $x$}{
		\BlankLine
			\eIf{distance($y$) = nil}{
				distance($y$) $\leftarrow$ distance($x$) + 1\;
				predecessor($y$) $\leftarrow$ $x$\;
				tail of QUEUE $\leftarrow$ $y$\;
				\BlankLine
				\eIf{($y$ is the only out-neighbour of $x$) AND ($x \neq i$)}{
					Add a Varc to the data structure using arcs $xy$ and the one formed by predecessor($x$) and $x$\;
				}{
					Add arc $xy$ to the data structure\;
				}
			}{
				\eIf{predecessor($y$) = nil}{
					Arc $xy$ is just a cycle back to the origin, ignore it\; 
				}{
					Add a chain to the data structure using (transitive) arc $xy$\;
				}
			}
		}
	}

	\caption{Modified BFS algorithm}
	\label{alg:main_algorithm}
\end{algorithm}

As illustrated in Algorithm \ref{alg:main_algorithm}, when a vertex is visited for the first time ($distance(y) = nil$) it is added to the data structure as either an arc, or as part of a Varc. Conversely, vertices that have been visited before may be included in a chain and need to be further processed to determine the length and vertices that could form such chain. The final objective of our program is to create and store the topological information of the announcement digraph in a Chain Routing data structure. Algorithm \ref{alg:main_algorithm} was implemented in a C++ program, which builds a Chain Routing data structure by finding most of the available chains and paths from vertex $i$ to the other 44 destinations (vertices). The specific functions that process the chains and implement the database are too large to be included in this paper.

%Although it is difficult to predict the behaviour of this program if the network is composed of thousands of ASs, it is expected that the program should be able to cope with the increase in size of the adjacency matrix. There are two reasons behind this assertion: This experiment was performed using the core of the European Internet which is the most connected part of the network; and the connectivity of this network is not as high outside this core.

The 45 announcement digraphs of the European ASs included in our previous study \cite{Arjona-Villicana2009} were processed and analyzed using the modified BFS algorithm. Then the longest possible chain or structure to each of the other 44 announced ASs was recorded in a large table ($45 \times 45$) which is available in \cite{Arjona-Villicana2010a}. Table \ref{tab:ChainsToOtherAS} shows the results for only the first 10 ASs in original original list of ASs. A description of what the entries in this table mean follows:

\begin{description}
	\item [number] This is the height of the longest chain between the source AS and the announced AS. When this number is 1, it indicates that there is only one arc between the ASs, but the announced AS is part of a longer chain. This means that, although there is no path diversity to this AS, it is still crucial to the connectivity of other ASs.
	\item [A] This entry means that there is just an \emph{arc} between the source AS and the announced AS. There is no further path diversity available.
	\item [B] This means that there is a \emph{bridge} between the source AS and the announced AS. This means that, although a chain is present in the path, at some point only an arc separates the two ASs. Therefore, a bridge indicates that the connectivity between the ASs is limited.
\end{description}

\begin{table*}
	\centering
	\scriptsize
	\renewcommand{\arraystretch}{1.3}
	\caption{Longest possible chain or structure from an originating AS to other ASs}
	\label{tab:ChainsToOtherAS}
	\begin{tabular*}{0.9\textwidth}{@{\extracolsep{\fill}} l c c c c c c c c c c }
		\hline	AS	&	AS1299	&	AS702	&	AS3303	&	AS1257	&	AS13237	&	AS8220	&	AS286	&	AS3257	&	AS1273	&	AS16150\\
		\hline	AS1299	&	-	&	1	&	3	&	A	&	A	&	A	&	2	&	A	&	2	&	B	\\
						AS702	&	2	&	-	&	2	&	2	&	3	&	3	&	3	&	3	&	2	&	3	\\
						AS3303	&	2	&	1	&	-	&	2	&	2	&	4	&	2	&	2	&	2	&	3	\\
		\hline	AS1257	&	2	&	1	&	2	&	-	&	2	&	A	&	2	&	A	&	2	&	B	\\
						AS13237	&	2	&	2	&	2	&	2	&	-	&	2	&	3	&	2	&	2	&	2	\\
						AS8220	&	2	&	1	&	2	&	A	&	2	&	-	&	2	&	1	&	2	&	2	\\
		\hline	AS286	&	2	&	1	&	2	&	A	&	2	&	2	&	-	&	A	&	1	&	2	\\
						AS3257	&	2	&	2	&	2	&	2	&	2	&	2	&	3	&	-	&	2	&	2	\\
						AS1273	&	2	&	2	&	2	&	2	&	2	&	2	&	3	&	2	&	-	&	2	\\
						AS16150	&	1	&	2	&	3	&	B	&	2	&	3	&	3	&	2	&	B	&	-	\\
		\hline
	\end{tabular*}
\end{table*}

\begin{table*}
	\centering
	\scriptsize
	\renewcommand{\arraystretch}{1.3}
	\caption{Number of arc-disjoint paths from an originating AS to other ASs}
	\label{tab:ArcDisjointPaths}
	\begin{tabular*}{0.9\textwidth}{@{\extracolsep{\fill}} l c c c c c c c c c c }
		\hline	AS	&	AS1299	&	AS702	&	AS3303	&	AS1257	&	AS13237	&	AS8220	&	AS286	&	AS3257	&	AS1273	&	AS16150\\
		\hline	AS1299	&	-	&	1	&	3	&	1	&	1	&	1	&	3	&	1	&	2	&	1	\\
						AS702	&	2	&	-	&	2	&	2	&	4	&	4	&	3	&	3	&	2	&	6	\\
						AS3303	&	2	&	1	&	-	&	3	&	3	&	5	&	4	&	3	&	2	&	4	\\
		\hline	AS1257	&	3	&	1	&	2	&	-	&	2	&	1	&	3	&	1	&	2	&	1	\\
						AS13237	&	2	&	2	&	4	&	3	&	-	&	3	&	3	&	3	&	3	&	3	\\
						AS8220	&	2	&	1	&	2	&	1	&	2	&	-	&	2	&	1	&	3	&	3	\\
		\hline	AS286	&	3	&	1	&	2	&	1	&	2	&	3	&	-	&	1	&	1	&	3	\\
						AS3257	&	2	&	2	&	4	&	3	&	3	&	3	&	4	&	-	&	2	&	2	\\
						AS1273	&	2	&	3	&	3	&	2	&	2	&	3	&	3	&	2	&	-	&	2	\\
						AS16150	&	2	&	3	&	4	&	2	&	3	&	4	&	4	&	2	&	2	&	-	\\
		\hline
	\end{tabular*}
\end{table*}

Table \ref{tab:ArcDisjointPaths} shows the number of arc-disjoint paths between the same set of ASs, but using the results obtained in \cite{Arjona-Villicana2009}.
By comparing Tables \ref{tab:ChainsToOtherAS}  and \ref{tab:ArcDisjointPaths}, it is possible to see that there is a strong relation between the number of arc-disjoint paths found and the height of the chains obtained by our program. This indicates that our basic implementation of the Chain Routing framework is efficient at exploiting the path diversity of a digraph.

Since the height of a chain and the number of arc-disjoint paths are equal, the results in Table \ref{tab:ChainsToOtherAS} show that in most cases there are 2 arc-disjoint paths to reach a destination, and that sometimes there is enough connectivity to build 3, or even 4 arc-disjoint paths to a destination, such as between AS3303 and AS8220. This demonstrates that Chain Routing can be employed to find and use alternative paths which may increase the resilience between a source and a destination.

On the other hand, we also noticed that some chains conflicted with each other. An example of such a conflcit would be $C_1(a, b, c)$ and $C_2(a, c, b)$. This situation implies that, if Chain Routing is going to exploit the connectivity between $a$, $b$, and $c$, it will need to select and use only one of $C_1$ or $C_2$, otherwise cyclic behavior could arise between these two structures. It also means that we may need to use a chain that is convenient as a general routing strategy, even if this is not the best solution for a particular destination.

The complete table, available in \cite{Arjona-Villicana2010a}, shows that some AS can only use bridges (B) and arcs (A) to reach the other 44 destinations. These ASs mostly rely on a better-connected AS to route to the rest of the network. This is probably what ASs without many communication links experience in the Internet. It also implies that ASs with limited connectivity will enjoy fewer benefits from the Chain Routing framework.

\begin{figure}[t]
	\centering
		\includegraphics[width=2.9in]{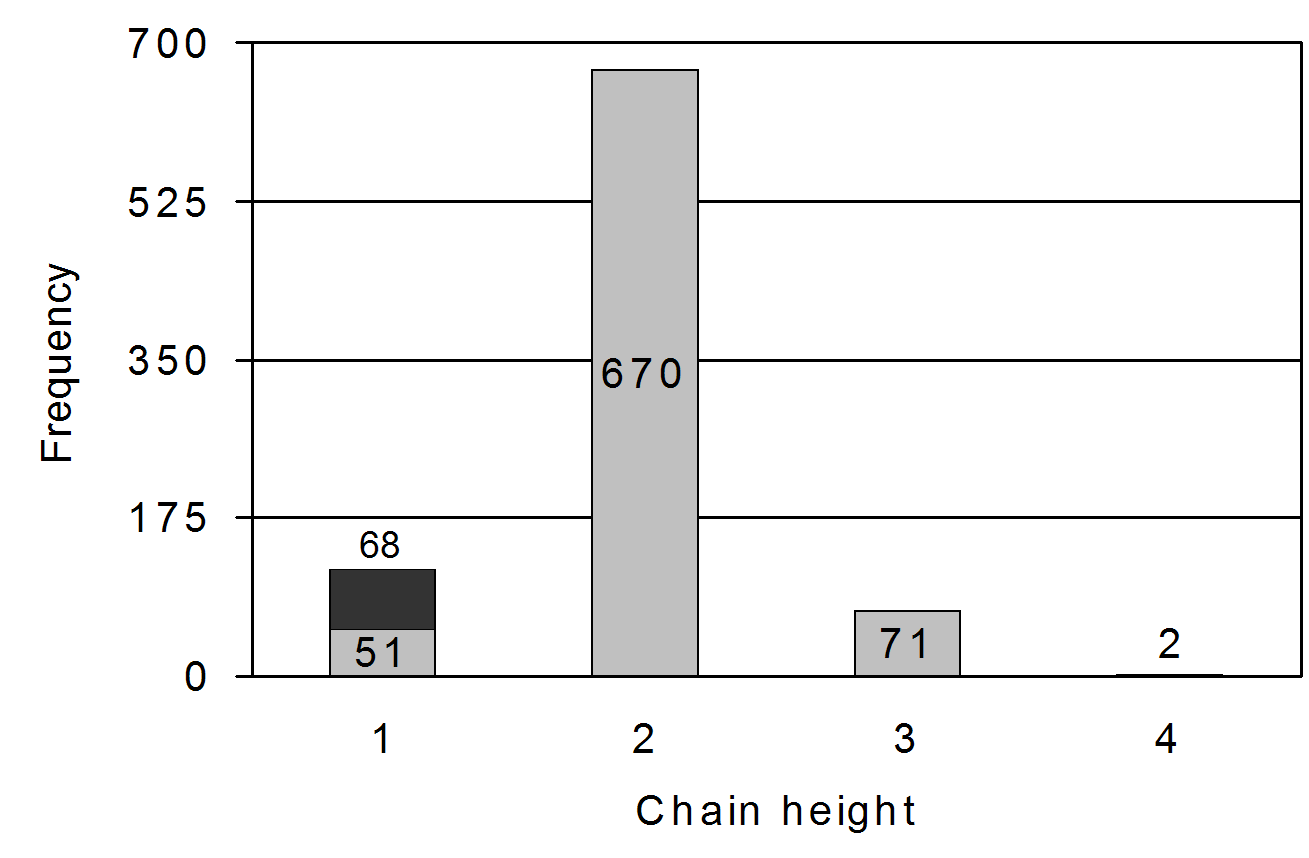}
	\caption{Frequency of chain height found}
	\label{fig:FreqofHeight}
\end{figure}

Fig.\ \ref{fig:FreqofHeight}  shows the frequency of each chain-height recorded in the complete table \cite{Arjona-Villicana2010a}. Since chains of height 1 and arcs (A) are similar, they are both counted under the same column (height = 1) where arcs appear in darker color. There were only 2 ocurrences of chains of height 4, and the most frequent chain, with a height of 2, had a count of 670. This figure does not show a number of entries which produced invalid results due to our program failing to process some announcement digraphs. These failures were caused by functionality that was still under development.

The results obtained from this analysis and the fact that the Chain Routing implementation used here is neither optimized nor complete, call for further development of software that can automatically search and define chains in the Internet. Such enhanced algorithm may implement either post-processes to combine chains and nested structures that were not initially discovered by our BFS algorithm, or functions specialized on discovering chains, instead of nodes.

%This is because the implemented program finishes after the modified BFS algorithm has completed its search for vertices in the digraph. However, after this algorithm has concluded, there is a possibility to combine chains and nested structures to define longer chains. Therefore, in order to take advantage of the full path diversity of a network, it is necessary to implement post-processes which can discover the partial orders that could not be found before, because full topological information was not available. There is also a need to explore if it is possible to apply or develop more efficient algorithms which would specialize on discovering chains, instead of nodes.

\section{Applying Chain Routing}\label{applying}

This section has two objectives. First, it analyzes how Chain Routing would increase network stability by demonstrating how this framework could help solve or ameliorate the effects of three of the most documented Internet pathologies (\ref{internet_stability}). Then, it considers the cost of fully implementing and employing Chain Routing (\ref{the_cost}) in a network.

\subsection{Chain Routing and Internet stability}\label{internet_stability}

\subsubsection{Persistent Route Oscillations}\label{is_pro}

Although PRO \cite{Varadhan1996} develop in networks that have announcement digraphs with cycles, this pathology also has a temporal aspect caused by the path selection process which is continously executing in BGP. This means that in order to avoid cyclic behaviors within a chain, it is necessary to consider the dynamics of this system. Therefore, we propose a pair of mechanisms which avoid the development of PRO in a chain. These have been converted into rules which can be implemented in a Chain Routing system:

\begin{Rule}\label{R:David01}
Before accepting to become part of a chain, every AS needs to verify that the proposed chain does not create a cycle with any other basic structure already defined in their own data structure.
\end{Rule}

\begin{figure}[!t]
	\centering
		\includegraphics[width=1.3in]{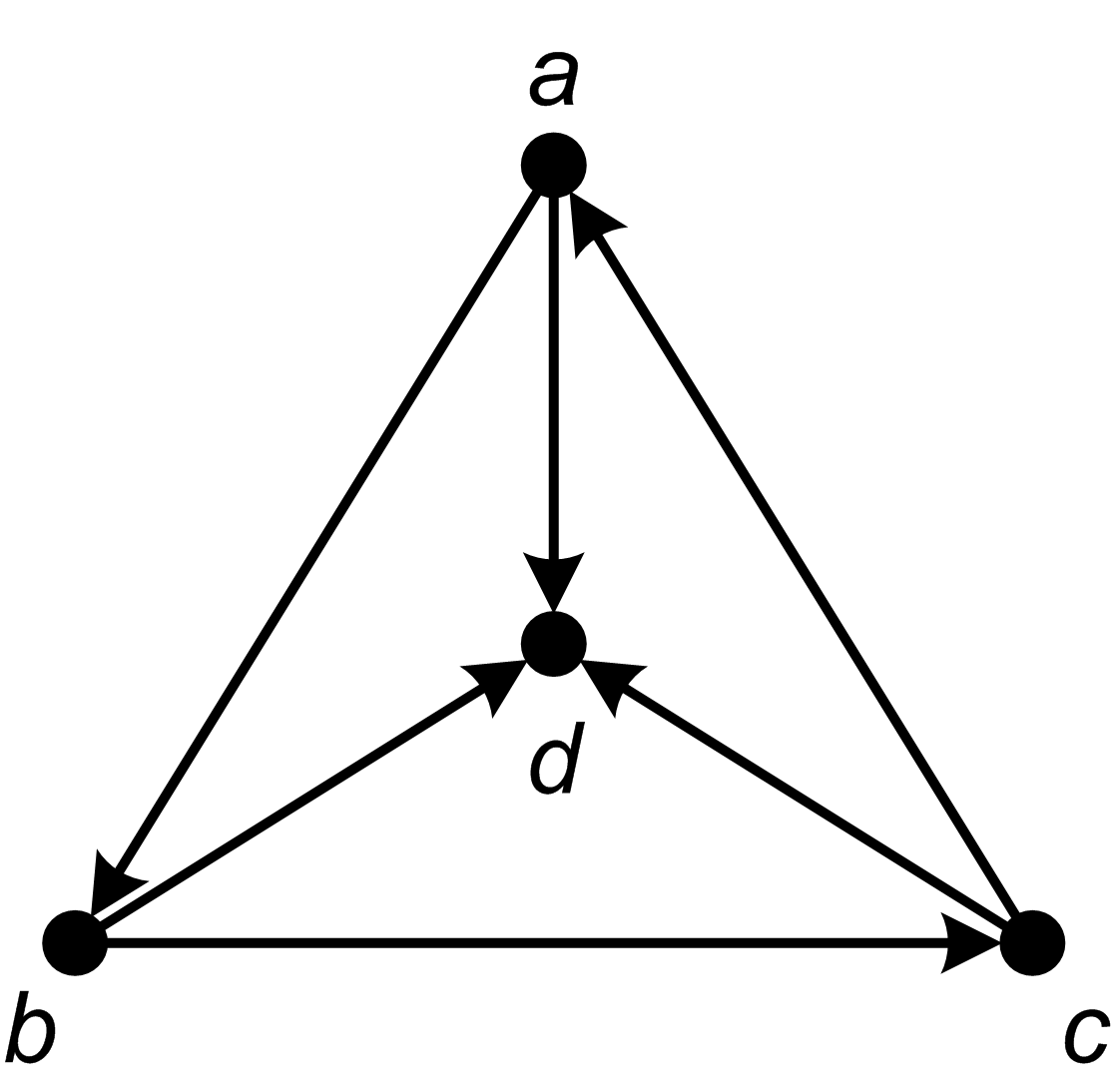}
	\caption{The cannonical PRO network by Varadhan et al.}
	\label{fig:71_Varadhan}
\end{figure}

This is similar to BGP's current functionality where ASs are constantly monitoring that cycles do not develop in their paths. To demonstrate how Rule \ref{R:David01} stops PROs from developing, we will use the cannonical PRO example first introduced by Varadhan et al.\ \cite{Varadhan1996} and depicted in Fig.\ \ref{fig:71_Varadhan}, in which ASs $a$, $b$ and $c$ have different options to reach AS $d$, but all of them prefer to use their longer path through the next neighbor over their shorter direct path. This preference is described in Table \ref{tab:PathPreferenceForTheClassicPRO} and it causes an endless sequence in which every AS prefers the path through their counter clockwise neighbor over their own shortest path. To demonstrate that this cyclic behavior will not develop if Rule \ref{R:David01} is applied, we provide the following example:

\begin{example}\label{E:example1}
Suppose that $C_a\left(a, b, d\right)$ is the first chain defined in the network depicted in Fig.\ \ref{fig:71_Varadhan}. This is a perfectly valid chain irrespective of which route AS $a$ uses to reach $d$ (either the direct $ad$ route or the one through $b$). Now suppose that $c$ tries to use $a$ to reach $d$, but because there is only one path from $c$ to $b$ ($ca+ab$), it will define chain $C_c\left(c, a, d\right)$ where the segment $ad$ is actually $C_a$. Finally, AS $b$ tries to use $c$ to reach $d$, but when $b$ requests to create chain $C_b\left(b, c, d\right)$, $c$ will apply Rule \ref{R:David01} and realize that $b$ is already part of $C_a$ (and $C_c$) and will reject creating $C_b$. Thus the PRO has been avoided.
\end{example}

\begin{table}[!t]
	\centering
  \footnotesize
	\renewcommand{\arraystretch}{1.3}
	\caption{Path preference for the cannonical PRO network}
	\label{tab:PathPreferenceForTheClassicPRO}
		\begin{tabular*}{0.4\textwidth}{@{\extracolsep{\fill}} c c c }
			\hline
			AS & Paths to $d$ & Preference\\
			\hline
			$a$ & $a \rightarrow b \rightarrow d$ & 1\\
					& $a \rightarrow d$   & 2\\
			\hline
			$b$ & $b \rightarrow c \rightarrow d$ & 1\\
					& $b \rightarrow d$   & 2\\
			\hline
			$c$ & $c \rightarrow a \rightarrow d$ & 1\\
					& $c \rightarrow d$   & 2\\
			\hline
		\end{tabular*}
\end{table}

The second rule to avoid PRO in a chain is:

\begin{Rule}\label{R:David02}
When a segment in a chain becomes unavailable and an alternative path needs to be used, it is safer to select paths that, because of the chain's topology, cannot route information through the unavailable segment.
\end{Rule}

Once it has been confirmed that paths, which are adjacent to the failed segment, can still reach the destination, they may be reinstated as safe paths.

\begin{figure}[!t]
	\centering
		\includegraphics[width=1.8in]{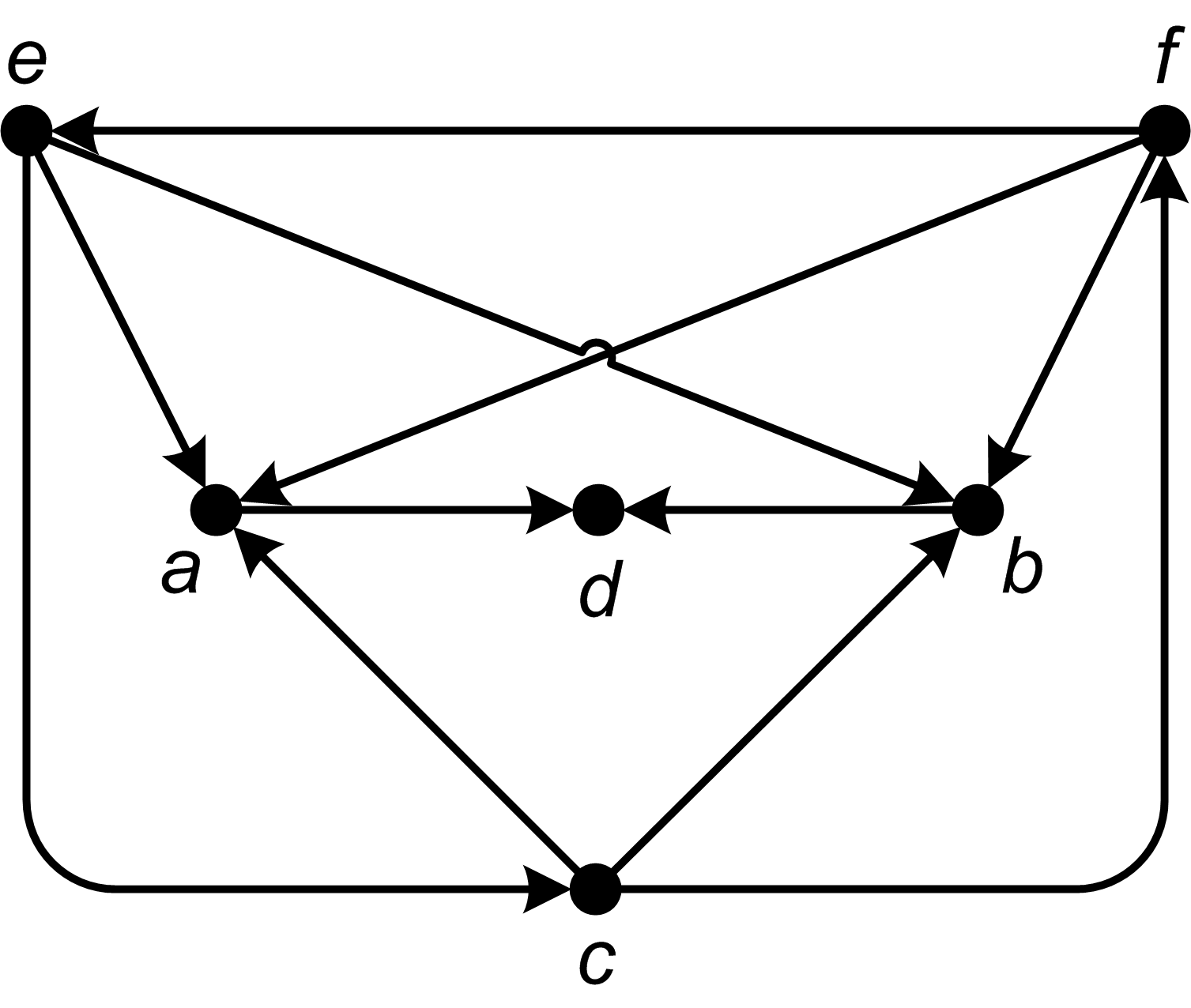}
	\caption{Another PRO case by Griffin et al.}
	\label{fig:72_Griffin}
\end{figure}

This type of behavior helps the network to quickly reach a stable state because alternative paths that have a probability of failure are not used. An example of how to apply this rule is illustrated in Fig.\ \ref{fig:72_Griffin}, which was originally introduced by Griffin et al.\ \cite{Griffin1999a}. The route preference for this network is described in Table \ref{tab:PathPreferenceForTheGriffinPRO}, where $X$ represents the fact that the originating AS's policy only requires that this alternative path sends information through its counter clockwise neighbor and finishes in $d$. Initially, ASs $c$, $e$ and $f$ will all send information to $d$ through AS $b$, but when link $bd$ fails, $c$, $e$ and $f$ will prefer to use their counter clockwise neighbor instead of the most direct route through $a$. This produces the same cyclic system and behavior described in Fig.\ \ref{fig:71_Varadhan}. The following example analyzes what happens when Chain Routing and Rule \ref{R:David02} are applied:

\begin{table}[!t]
	\centering
  \footnotesize
	\renewcommand{\arraystretch}{1.5}
	\caption{Path preference for the PRO network by Griffin et al.}
	\label{tab:PathPreferenceForTheGriffinPRO}
		\begin{tabular*}{0.4\textwidth}{@{\extracolsep{\fill}} c c c }
			\hline
			AS & Paths to $d$ & Preference\\
			\hline
					& c $\rightarrow b \rightarrow d$   & 1\\
			$c$ & c $\rightarrow f \rightarrow X \rightarrow d$ & 2\\
					& c $\rightarrow a \rightarrow d$   & 3\\
			\hline
					& e $\rightarrow b \rightarrow d$   & 1\\
			$e$ & e $\rightarrow c \rightarrow X \rightarrow d$ & 2\\
					& e $\rightarrow a \rightarrow d$   & 3\\
			\hline
					& f $\rightarrow b \rightarrow d$   & 1\\
			$f$ & f $\rightarrow e \rightarrow X \rightarrow d$ & 2\\
					& f $\rightarrow a \rightarrow d$   & 3\\
			\hline
		\end{tabular*}
\end{table}

\begin{example}
In the network depicted in Fig.\ \ref{fig:72_Griffin}, a Chain Routing system could define the following three chains:

\begin{enumerate}
	\item $C_1\left(c, f, b, d\right)$ with Varcs $ca+ad$ and $fa+ad$.
	\item $C_2\left(e, c, b, d\right)$ with Varcs $ea+ad$ and $ca+ad$.
	\item $C_3\left(f, e, b, d\right)$ with Varcs $fa+ad$ and $ea+ad$.
\end{enumerate}

So when link $bd$ fails, ASs $c$, $e$ and $f$ will comply with Rule \ref{R:David02} and select paths that, because of the chain's topology, cannot route information through the faulty $bd$ segment. In this case, the direct path of each chain would be the only route that agrees with Rule \ref{R:David02}. For example, AS $c$ would select the segment $cd$ (Varc $ca+ad$) which cannot route information through segment $bd$.
\end{example}

By assuring that no cycles develop when the topology of a network changes, these rules provide Chain Routing with enough robustness to avoid PROs in a network.

\subsubsection{Delayed network convergence}

We explained in Section \ref{conv_delays} that path exploration in a BGP system may cause transient loops and dropped packets. Chain Routing uses complete orders as its topological unit, which are acyclic digraphs. Therefore, it is possible to guarantee that transient loops will not develop. Conversely, because Chain Routing does not control the dynamics of the system, it cannot assure that the network will reach faster convergence times, nor that information packets will not get lost while the network is in its transient state. However, in Section \ref{comp_orders_in_time} we will illustrate that by combining complete orders in time and topology it may be possible to shorten the duration of transient instabilities in the Internet.

\subsubsection{Network congestion}

Chain Routing provides a framework which can be employed to perform traffic engineering between all its arc-disjoint paths. In general, by applying Theorem \ref{T:Num2}, a chain of $n$ vertices could use $n-1$ arc-disjoint paths to distribute traffic between the source and destination. Therefore, although Chain Routing cannot directly eliminate congestion in a network, it allows to implement better traffic administration mechanisms to avoid this problem.

In conclusion, the Chain Routing framework may help to increase network resilience at the same time that exploits the Internet's path diversity, but it cannot solve all the instabilities observed in the Internet.

\subsection{The cost of implementing Chain Routing}\label{the_cost}

%In this section we will consider the costs of implementing and employing Chain Routing in a network.

Chain Routing propagates its routing information via announcement digraphs, which possess more arcs and better connectivity than the BGP digraph \cite{Arjona-Villicana2010}. Unfortunately, the increased connectivity of the announcement digraph also requires more control messages to define chains. These extra-messages depend on the policies that each AS applies to its neighbors; therefore it is not possible to accurately predict the exact number needed to define such chains in a network. Besides the additional messages produced by using the announcement digraph, it is necessary to consider that coordination messages will be needed between the ASs that form chains in order to mantain these structures. Although the details of the messages needed to establish such chains are outside the scope of this article, a tentative solution would need:

\begin{enumerate}
	\item A message from the source to every intermediate ($n-2$) node requesting to establish a chain.
	\item A reply message from every intermediate node accepting or rejecting to be part of the chain.
	\item Another message from the source to the intermediate nodes that have accepted to be part of the chain confirming that the chain has been implemented.
\end{enumerate}

%Still, the numerical analysis described in this chapter suggests that many more destination messages would be travelling in the network if the announcement digraph is used, without employing the PPR.

%Another important consideration is the number of extra-messages needed to establish a chain. In order to discover the available alternative paths to reach a destination it suffices to use the announcement digraph. As it has been shown in section \ref{mt_bgp_policies_create_a_digraph}, the announcement digraph has more links than the currently used BGP digraph; the difference between these two digraphs will depend on the policy restrictions that each AS in the network will impose and hence, it is not possible to quantify it accurately. Still the experiment in section \ref{cr_implementing_chain_routing_in_c} suggests that many more destination messages would be traveling in the network if the announcement digraph is used.

This means that every chain of length $n$ may need up to $3(n-2)$ extra-messages to establish each chain. In practice, there may be conflicts when deciding who would be the source and the order of the intermediate vertices, and this may cause the number of messages to increase while a suitable chain is defined.

It is also necessary to consider the amount of resources required to store and manage the data structure that will record chains. Although this data structure will contain a list of chains to different destinations, more than one destination could be recorded in each chain, therefore Chain Routing should employ a smaller number of chains than the number of destinations in the network. Since this paper only provides an outline of the Chain Routing framework, we are not calculating this cost here and will address it as part of future research. Nevertheless, it is important to remember that the increased cost of managing and storing the Chain Routing data structure is what enables to keep more topological information of the network.

\section{Beyond Topology: Complete Orders in Time}\label{comp_orders_in_time}
 
We recognize that any routing protocol not only needs to find the best paths to reach a destination, but also to adapt to sudden changes in topology. Therefore, there is a necessity to consider time as another important constraint in the behavior of a routing protocol. Subsection \ref{temp_ord_topo_ord} analyzes the interactions between events that follow a temporal order or a topological order, and demonstrates that when either of these orders fail to exist instabilities may arise in the network. Then, subsection \ref{timescales} proposes to use different timescales to allow Chain Routing to mantain the stability of the network.

\subsection{Temporal order vs. topological order}\label{temp_ord_topo_ord}

The following example demonstrates what could happen when there is no clear order in the time realm.

\begin{figure}[!t]
	\centering
		\includegraphics[width=2.4in]{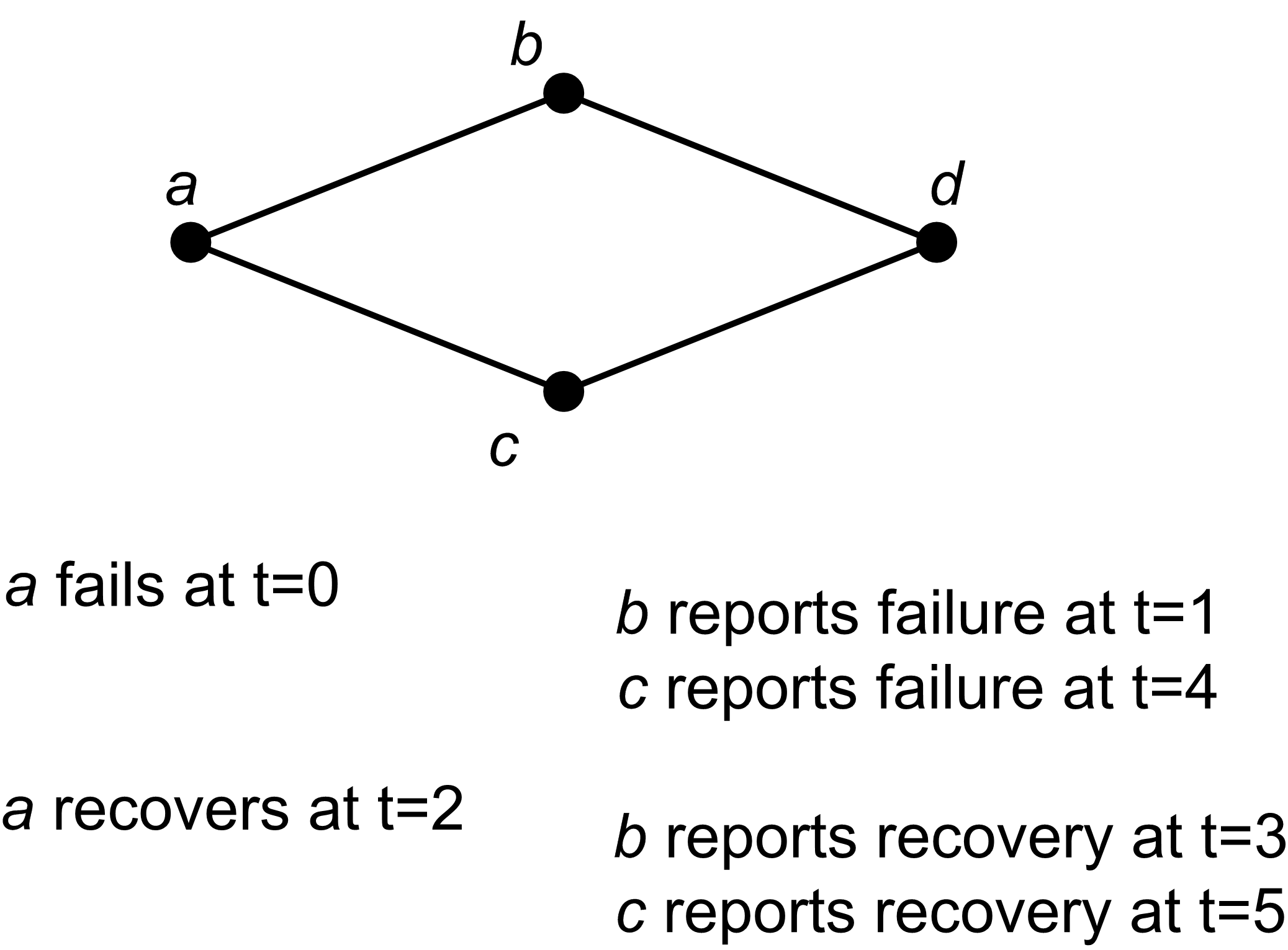}
	\caption{Network with no termporal order}
	\label{fig:73_TempOrder}
\end{figure}

\begin{example}
In the network depicted in Fig.\ \ref{fig:73_TempOrder}, AS $a$ suffers a failure at $t=0$, but recovers from it at $t=2$. AS $b$ has a fast link to $a$ and $d$, therefore $b$ quickly reports $a$'s failure and recovery. On the other hand, $c$ cannot pass routing information to $d$ as fast as $b$ does. This causes a delay in the messages which corrupts the order at which they arrive at $d$. Therefore, $d$ sees that $a$ failed at $t=1$, recovered at $t=3$, failed again at $t=4$ and recovered at $t=5$. This lack of order in time may also cause other problems in larger networks \cite{Mao2002}.
\end{example}

In order to avoid the route flapping presented in the previous example, it is necessary to implement an order in time. A simple way to achieve this is by reporting not only the occurrence of an event, but also the time when it happened \cite{Chandrashekar2005, Pei2005}. If the messages in Fig.\ \ref{fig:73_TempOrder} include the time at which the failure ($t=0$) and the recovery ($t=2$) happened, when AS $d$ receives the delayed failure message from $c$ at $t=4$, $d$ would be able to determine that this is an event that happened before the recovery message announced by $b$ at $t=3$.

In contrast, the following example shows a network in which there is no clear order in topology.

\begin{figure}[!t]
	\centering
		\includegraphics[width=2.2in]{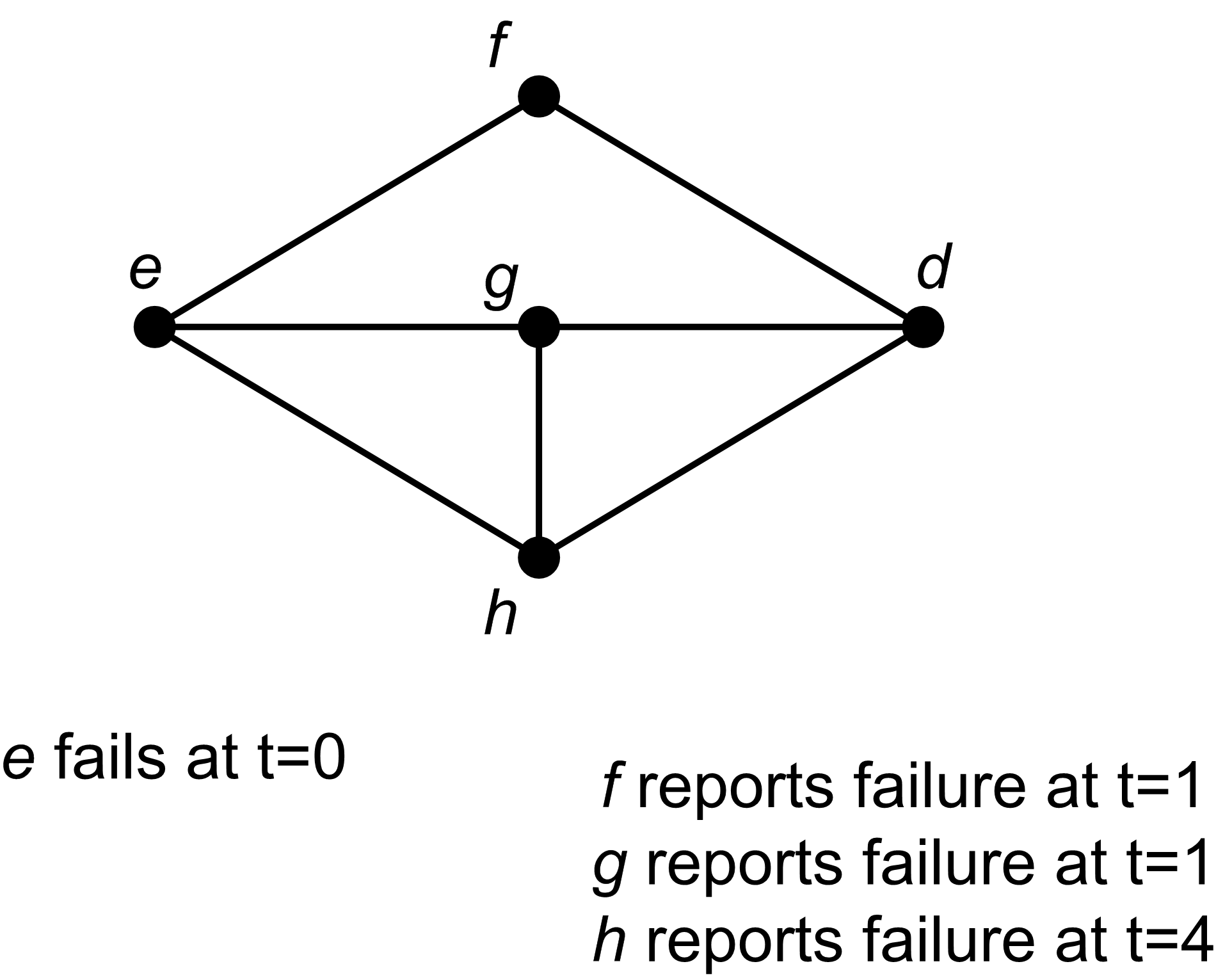}
	\caption{Network with no topological order}
	\label{fig:74_TopoOrder}
\end{figure}

\begin{example}
In the network shown in Fig.\ \ref{fig:74_TopoOrder}, AS $e$ fails at $t=0$, and ASs $g$ and $f$ acknowledge this failure at $t=1$, but $h$ takes more time to detect that $e$ has become unavailable; this causes $g$ and $d$ to insist that they can still route data through $h$. It is not until $t=4$ that $h$ reports that it cannot reach $e$, and then $g$ and $d$ stop trying to send data through $h$.
\end{example}

The instability shown in the previous example could have been avoided if an order in topology is established before the failure occurrs. For example, if chain $C(d, h, g, e)$ is defined, when $e$ fails and $g$ reports the failure at $t=1$, because $h$ is lower than $g$ in its Hasse diagram, the vertex order will forbid $g$ to use $h$ to reach $e$, thus the problem is avoided. On the other hand, if the chain $C(d, g, h, e)$ is defined, Rule \ref{R:David02} (see \ref{is_pro}) will force AS $d$ to use its direct link $Varc(df+fe)$ to reach $e$, but because $f$ also reported the failure at $t=1$, $d$ will be aware that $e$ has become unavailable.

In Fig.\ \ref{fig:74_TopoOrder}, if we route to $e$ using $C(d, h, g, e)$, $h$ has two paths to the destination ($h \rightarrow e$ and $h \rightarrow g \rightarrow e$), while $g$ has only one ($g \rightarrow e$). Conversely, if we use  $C(d, g, h, e)$, $h$ has only one path to $e$ and $g$ has two. This means that vertices that are lower in the Hasse diagram of a chain have more paths and better connectivity to the destination than those that are higher.

The previous two examples demonstrate that by combining orders in time and topology it is possible to maintain network stability, but when either of these two orders is absent, instablities will occur in the network due to inaccurate information.

\subsection{Chain Routing timescales}\label{timescales}

As it was asserted in Section \ref{the_cost}, Chain Routing will need at least three rounds of messages to establish a chain. However, once a chain has been defined all ASs will be communicated to every other AS via a chain segment. Therefore, we will define two timescales for the Chain Routing framework: the \emph{long-term scale} will be similar to the period of time needed to define a chain, while the \emph{immediate timescale} will be comparable to the time needed to transmit a message between two ASs in a chain. These timescales would help to support a stable and reliable connectivity where only necessary changes are sudden. Hence, at the immediate timescale, an AS should use a selected chain to a destination for as long as there is a viable path in that structure, and only if no path is available, the source will switch to a different chain. Meanwhile, at the long-term scale, the source should constantly be monitoring for a longer or more stable chain to reach the same destination, and only when all the involved ASs have reached an agreement, the source will be free to switch and use the new chain.

It is also important to consider that, just like BGP becomes unstable when it cannot find a suitable set of paths to reach a destination, it is also possible that Chain Routing cannot find a suitable set of chains in a network. This may produce oscillations between competing alternative chains. Fortunately, this is a different type of problem in which it is possible to stop the oscillations without affecting the traffic because every chain is a robust solution. Still, it is necessary to develop mechanisms that eliminate the possibility of Chain Routing becoming unstable because it cannot find a definitive chain to reach a destination.

\section{Discussion}\label{discusion}

As a routing framework, Chain Routing offers many advantages over BGP's current implementation. The main one is that failure of a link or an AS does not require the protocol to reconverge, because each AS knows beforehand alternative paths to reach a destination. This is the result of Chain Routing's topological unit, the complete order, and its two timescales, immediate and long-term, which allow it to define more than one arc-disjoint path to a destination.

Other advantages of employing the Chain Routing framework are that:

\begin{enumerate}
	\item it allows easy implementation of traffic engineering. This is because, once a chain has been established, it is trivial to identify the other stable paths and to use them concurrently.
	\item failures will not cause transient data loops in the network.
	\item it is easier to identify and prevent oscillatory behaviors (PRO).
\end{enumerate}

Conversely, Chain Routing will produce some challenges that need to be addressed before a final implementation is proposed. A problem that derives from the increased complexity of Chain Routing is that route aggregation may be difficult once chains to destinations have been established. This happens because a new set of IP addresses would require further network coordination which may result in an intermediate AS rejecting the modifications and impacting a previously defined chain.

Also, since vertices lower in the Hasse diagram of a chain will enjoy more connectivity than those at the top, there might be disputes on the order that ASs will follow when defining a chain. This means that, in some cases, human intervention and negotiation may be required to form chains. It also means that the traditional customer-provider model \cite{Gao2000a} might need to be reconsidered and perhaps superseeded by another economic model which accommodates for chains.

Another factor that has not been addressed by this research is the potential interactions between interior gateway protocols and Chain Routing. It has been previously demonstrated that such interactions are sometiems problematic for BGP \cite{Labovitz1999}. Therefore, it makes sense to develop a solution that allows safe and stable interactions between different types of routing protocols.

\section{Conclusion}\label{conclusion}

In this paper we have proposed the development of a routing framework whose main topological unit is the complete order: Chain Routing. Such framework allows to exploit the Internet's unused path diversity while, at the same time, maintains the stability of this network. The main advantages of using complete orders are that: it allows easy implementation of traffic engineering, enables the nodes in the chain to quickly find alternative paths when a failure occurs and it avoids the occurrence of transient loops. Although Chain Routing is a more stable solution than the current BGP implementation, it also requires more coordination between ASs.

Still, this proposed framework is just a theoretical solution that requires further empirical development and testing. This research has only laid down the foundations of a new routing scheme, and its final implementation was not included in the scope of this paper. There are also many characteristics that were not sufficiently addressed in here, such as the influence and implementation of policies, the fast growth and shrinkage of chains, the influence of Chain Routing in the economic model of the network and the mechanisms which would allow route aggregation and scalability. Moreover, a large amount of experimentation is needed before a practical implementation of Chain Routing is obtained, which includes finding an efficient algorithm for discovering complete orders in a digraph.

%Up to this point we have described how to use Chain Routing to build acyclic structures that encode the path diversity available between $s$ and $d$. How to employ this path diversity is the same as determining how to route information from $s$ to $d$. Since the main objective of this article is not to fully define a routing protocol this topic will need to be completed by future research.

The conditions under which a network maintains its stability were also explored and it was determined that by applying complete orders in two realms, temporal and topological, it might be possible to obtain a highly stable routing protocol which is more resilient to failures than BGP's current implementation. This lends support to the assertion that the topological complete order presented here, Chain Routing, has the potential to become a very stable solution for routing in the Internet.

Finally, we believe it is possible that the application of Chain Routing could be extended to other systems which can be modelled as a digraph that needs to maintain its connectivity. Examples of such systems might be overlay networks, data center networks or even vehicular traffic distribution.

%---------------------------------------------------------------
\bibliographystyle{model1-num-names}%{plain}%{elsarticle-num}
\bibliography{CR_references}

\end{document}